# Two-dimensional Iron Monocarbide with Planar Hypercoordinate Iron and Carbon

*Dong Fan, Shaohua Lu, and Xiaojun Hu*

College of Materials Science and Engineering, Zhejiang University of Technology, Hangzhou 310014, China

We report on the theoretical discovery of Iron monocarbide binary sheets stabilized at two-dimensional confined space, which we call tetragonal-FeC (t-FeC) and orthorhombic-FeC (o-FeC), respectively. From the energy viewpoint, the proposed t-FeC is the global minimum configuration in the 2D space, and each carbon atom is four-coordinated with ambient four Iron atoms. Strikingly, the o-FeC monolayer is an orthorhombic phase with planar pentacoordinate carbon moiety and planar seven-coordinate Fe moiety. To our knowledge, this monolayer is the first example of a simultaneously pentacoordinate carbon and planar seven-coordinate Fe-containing material. State-of-the-art theoretical calculations confirm that all these monolayers have significantly dynamic, mechanical, and thermal stabilities. Among these two monolayers, t-FeC monolayer shows a higher theoretical capacity (395 mAh g$^{-1}$), and can stably adsorb Li up to t-FeCLi$_3$. Low migration energy barrier is predicted as small as 0.26 eV for Li, which result in the fast diffusion of Li atom on this monolayer. Moreover, electron-phonon calculations coupled with Bardeen-Cooper-Schrieffer arguments suggest t-FeC can be potential two-dimensional superconductors with 6.77 K superconducting transition temperature.

## 1. INTRODUCTION

As an amazing and versatile element, carbon has typically $sp^3$-, (*e.g.*, diamond) $sp^2$-, (*e.g.*, graphene)[1] or linear *sp*-hybridized (*e.g.*, carbyne)[2] bonding networks. Especially, since the planar tetracoordinate carbon (ptC) has been first proposed by Hoffman *et al.* (1970)[3] and later synthesized in metal compound molecules (1977),[4] numerous stable ptCs molecules have been demonstrated theoretically and experimentally, such as (2,6-dimethoxyphenyl)lithium,[5] C-B-H systems,[6] C-Al systems,[7] and B-C systems,[8,9] *etc*. Moreover, molecules containing planar carbon with even higher coordination, have also been reported, *e.g.*, Wang *et al.* theoretically predicted numerous minima molecules with planar hexacoordinate, heptacoordinate, and pentacoordinate carbons in B-C systems;[10] Aromatic boron wheels with hypercoordinate carbon atom in $C_2B_8$, $C_3B_9^{3+}$, and $C_5B_{11}^+$ molecules were proposed by Erhardt *et al.*;[11] Ito *et al.* reported a planar hexacoordinate carbon molecule, $CB_6^{2-}$, can be annulated by inserting arenes or olefins into a perimeter B-B bond.[12]

In recent years, there has been growing interest in the design of planar hypercoordinate carbon in the periodic systems planar hypercoordinate carbon, and the concept of planar hypercoordinate compounds also be expanded to other elements.[13,14] $B_2C$ graphene structure containing the ptC moiety and planar $Be_2C$ monolayer with quasi-planar hexacoordinate carbons atoms was proposed by Wu *et al.* and Li *et al.*,[13,15] respectively. Recently, Yang *et al.* reported the $Cu_2Si$ monolayer featuring planar hexacoordinate silicon and planar hexacoordinate copper, and this monolayer has been synthesized very recently.[16,17] However, there is rare literature precedent to date about planar hypercoordinate motifs in the extended carbon-transition metal systems.

In this work, a systematic optimization algorithm is performed to search stable two-dimensional (2D) materials containing planar hypercoordinate transition metal and carbon atoms. We selected Fe as the typical representative of the transition metal, because it is abundant in the earth and its planar hypercoordination was demonstrated in the Fe-B systems.[18,19] By means of comprehensive density functional theory (DFT) computations, we identified two novels 2D materials, tetragonal-FeC (t-FeC) and orthorhombic-FeC (o-FeC), namely. For t-FeC monolayer, this novel structure can be considered as the quasi-planar tetragonal lattice, in which each C atom is quasi-plane four-coordinated with ambient four Fe atoms. In o-FeC monolayer, each C atom binds to five Fe and C atoms to form a perfect planar pentacoordinate moiety. Our results revealed that t-FeC monolayer can provide lower Li adsorption energy, faster Li mobility, and higher theoretical capacity than graphite and $MoS_2$. Combining such advanced features, t-FeC monolayer is a promising anode material for lithium ion batteries. We also demonstrate that t-FeC is a superconductor with a moderate electron-phonon coupling λ = 0.74, leading to $T_c \approx$ 6.77 K. By applying slightly uniaxial compressive strain, we are essentially able to enhance the $T_c$ up to 7.34 K. All these fascinating properties make t-FeC and o-FeC monolayers the promising candidates for future applications in nanoscale electronics.

## 2. COMPUTATIONAL DETAILS

Candidate structures were obtained by using a particle swarm optimization (PSO) method, as implemented in CALYPSO code,[20,21] it has been successfully applied to various crystal surfaces and low dimensional materials.[22–24] The vacuum space of 18 Å in the non-periodic direction was used to minimize the interaction between the neighboring layers. The subsequent structural relaxation and total energy calculations were carried out in the Vienna *ab initio* simulation package (VASP).[25] The electron exchange-correlation functional was treated by the Perdew−Burke−Ernzerhof (PBE) functional within the generalized gradient approximation (GGA) scheme.[26] The energy cutoff of the plane wave was set to 650 eV with the energy precision of $10^{-5}$ eV. The atomic positions were fully relaxed until the maximum force on each atom was less than $10^{-3}$ eV/Å. The Brillouin zone was sampled with a $12 \times 12 \times 1$ ($12 \times 18 \times 1$) Monkhorst-Pack k-points grid for static calculations. To account for strong correlation of unfilled d orbital of Fe atom, we also apply the GGA+$U$ scheme.[27] We carried out band structure calculations with an effective $U$ value of 5 eV, in keeping with previous studies.[28,29] Phonon dispersions and frequency densities of states (DOS) were performed in the Phonopy package[30] interfaced with the density functional perturbation theory (DFPT)[31] as performed in VASP. The finite temperature first principles molecular dynamics (FPMD) simulations were performed to further examine the stability of the structure by using time steps of 1 femtosecond in $3 \times 3 \times 1$ super-cells. The diffusion energy barrier and minimum energy pathway of Li diffusion on the t-FeC monolayer was calculated by using the nudged elastic band (NEB) method.[32]

Considering the charging/discharging processes of t-FeC monolayer anode can be assumed as: t-FeC + $x$Li$^+$ + $x$e$^-$ ↔ t-FeCLi$_x$, thus, the average open circuit voltage (OCV) for Li intercalation on the t-FeC monolayer was estimated by (volume and entropy effects are neglected):

OCV ≈ $[E_{FeCLix1} - E_{FeCLix2} + (x_2 - x_1)E_{Li}]/[(x_2 - x_1)e]$

where $E_{FeCLix1}$, $E_{FeCLix2}$ and $E_{Li}$ represent the total energy of t-FeCLi$_{x1}$, t-FeCLi$_{x2}$ and metallic Li, respectively.

The QuantumEspresso 6.1 package[33] was used to calculate the superconducting property. We used norm-conserving pseudopotentials (FHI) and a cutoff energy of 120 Ry. the Allen-Dynes modified McMillan formula[34] was used to estimate the superconducting temperature ($T_c$):

$$T_c = \frac{\omega_{log}}{1.2}\exp(-\frac{1.04[1 + \lambda]}{\lambda - \mu^*[1 + 0.62\lambda]})$$

where $\mu^*$ is the Coulomb pseudopotential, $\lambda$ is the overall electron-phonon coupling constant calculated from the frequency-dependent Eliashberg spectral function $\alpha^2 F_{(\omega)}$, and $\omega_{log}$ is the logarithmic average phonon frequency. A $4 \times 4 \times 1$ q-mesh was used together with a denser k-mesh for the 2D tetragonal FeC monolayer, resulting in a well-converged superconducting transition temperature $T_c$.

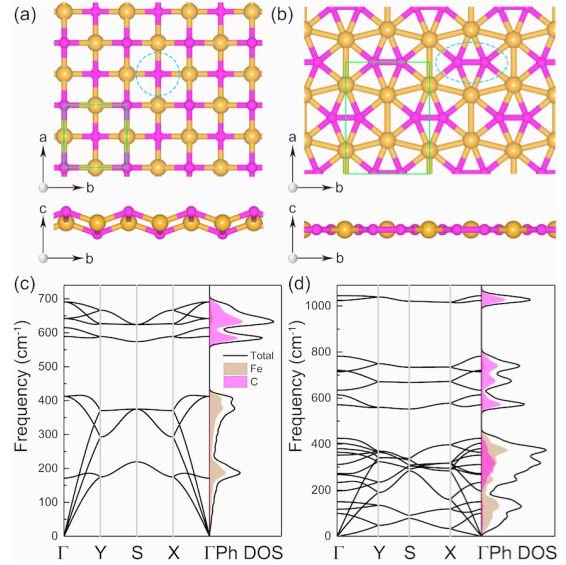

**Figure 1**. Top and side views of the proposed (a) t-FeC and (b) o-FeC monolayer, respectively. The green lines denote a unit cell; a and b represent the lattice vectors; the orange and magenta spheres refer to Fe and C atoms, respectively. Calculated phonon dispersion curves of (c) t-FeC and (d) o-FeC monolayer, respectively.

## 3. RESULTS AND DISCUSSION

The ground state structures of Fe-C compounds with 1:1 stoichiometry was selected through a comprehensive PSO search. We focus our discussion in this work on these two extraordinary phases, t-FeC and o-FeC, leaving a detailed analysis of all other predicted structures to a follow-up work.

As shown in Figure 1a, the relaxed t-FeC monolayer crystallizes in the tetragonal lattice with space group *P4/nmm*. This t-FeC sheet has a puckered square structure with a thickness $l$ = 1.26 Å. The calculated lattice constants are **a** = **b** = 3.49 Å, which is analogous to previously proposed 2D t-TiC,[35] t-YN,[36] t-SiC,[23] rectangular TiN,[37] and rectangular NbN.[38] Figure 1b presents the optimized structure of the designed o-FeC monolayer. One unit cell of o-FeC monolayer consists of four Fe and C atoms, with the optimized lattice constants of **a** = 5.96 Å, **b** = 4.49 Å, respectively. The length of Fe-C bond in this monolayer (1.85 Å) is noticeably smaller than that in bulk FeC$_2$ sheet (1.84 − 2.11 Å), indicating a much stronger interaction between the Fe and C atoms. Additionally, we also considered the buckled structure of o-FeC monolayer, and found that the planar structure is lower in energy than the buckled structure. Therefore, unlike the t-FeC monolayer with puckered structure, o-FeC monolayer shows a perfect planar lattice with space group *Cmmm*.

In this planar sheet, each C atom coordinates with five adjacent atoms, (one C atom and four Fe atoms) thus forming planar pentacoordinate moiety; while each Fe atom coordinates with four carbon atoms and three Fe atoms, forming a planar seven-coordinate moiety. The optimized Fe-C bond length (1.91 − 1.94 Å) is about the same as that in the $FeC_2$ sheet (1.84 − 2.11 Å), but slightly larger than that in the t-FeC monolayer (1.86 Å); while we notice that the bond length of Fe-Fe is 2.31 − 2.44 Å, which is close to previous results of 2.33 Å in the square iron membranes.[39,40]

To assess the energetic stability of the proposed structures, we first compare the relative energy of t-FeC and o-FeC structures with previously reported structures: square-FeC, and honeycomb-FeC, namely.[39] As shown in Table S1 and Figure S1, all functionals prefer t-FeC and o-FeC significantly compared with previously reported planar honeycomb and square lattices. We then calculated the cohesive energy to further chick the stability of proposed structures, which is defined as: $E_{coh} = (nE_{Fe} + mE_C − E_{Total})/(n + m)$, in which $E_{Fe}$, $E_C$, and $E_{Total}$ are the total energies of a single Fe atom, a single C atom and proposed monolayers, respectively; $n$ and $m$ are the numbers of Fe and C atoms in the unit cell, respectively. The calculated cohesive energies of t-FeC and o-FeC are 5.76 and 5.59 eV per atom, respectively, which are higher than those of $Be_5C_2$ (4.58 eV/atom),[41] $Be_2C$ (4.86 eV/atom),[13] $FeB_2$ (4.87 eV/atom),[18] and $Cu_2Si$ monolayer (3.46 eV/atom)[16] at the same theoretical level. Therefore, the even higher cohesive energy can ensure that t-FeC and o-FeC monolayers are strongly connected networks.

The stability of t-FeC and o-FeC monolayers was further investigated by calculating the phonon dispersion curves and phonon density of states (Ph DOS) along the high-symmetry directions. As shown in Figure 1c and 1d, there is no sign of an imaginary phonon frequency observed in the entire Brillouin zone. Simultaneously, the calculated Ph DOS also reveal that there is no phonon with imaginary frequency in these monolayers, which very well agree with the phonon dispersion curves. The results demonstrate that these two monolayers are dynamically stable. In particular, for o-FeC, the highest frequency of optical modes reaches around 1047 $cm^{-1}$, which is much higher than the highest frequency of 854 $cm^{-1}$ in $FeB_2$,[18] 580 $cm^{-1}$ in silicene,[42] and 420 $cm^{-1}$ in $Cu_2Si$ monolayer,[16] suggesting a robust Fe-C interaction. Therefore, although the calculated relative energy of o-FeC is higher than that of t-FeC by 0.34 eV per formula, the former still represents a strongly bonded network.

Moreover, to verify that the proposed new materials will be stable at ambient temperatures, we performed FPMD simulations using a 3 × 3 × 1 supercell for t-FeC and 2 × 3 × 1 supercell for o-FeC monolayers at different temperatures of 500, 1500, and 2000 K with a time step of 1 fs. After FPMD simulation of 5 ps, snapshots taken at the end of each simulation time are illustrated in Figure 2. From the snapshots, all proposed monolayers can maintain its structural integrity at 1500 K, and unexpectedly, t-FeC sheet shows high structural stability even at 2000 K. The above results reveal that the monolayers exhibit very high thermal stability and can maintain its structural integrity up to 1500 K.

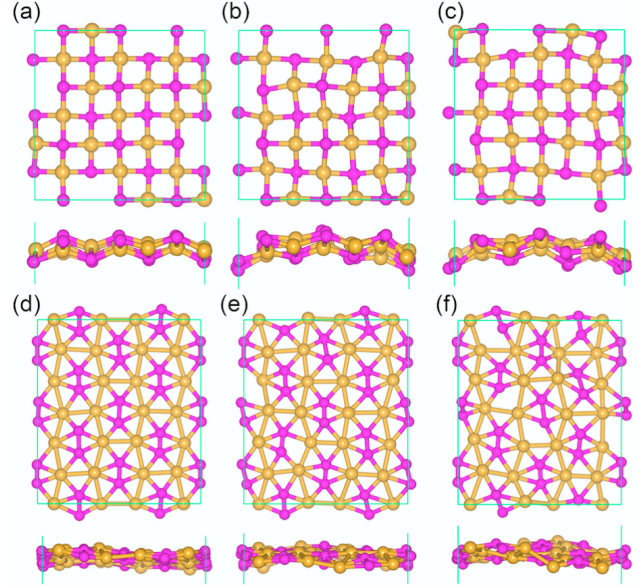

**Figure 2**. Snapshots of the (a, b, c) t-FeC and (d, e, f) o-FeC equilibrium structures at (a, d) 500 K, (b, e) 1500 K, (c, f) and 2000 K at the end of 5 ps FPMD simulations.

The mechanical properties are another important parameter for the potential applications of 2D materials. Generally, the in-plane stiffness, (or in-plane Young modulus), is used to assess the mechanical stability of 2D materials. As listed in Table S2, we compare the calculated value of the t-FeC and o-FeC sheets to previous experimental or theoretical values of several 2D materials, including graphene, borophone, and $B_2C$ sheets.[15,43] For the t-FeC monolayer, the in-plane stiffness was computed to be 77 N $m^{-1}$, which is lower than that of graphene. However, it is comparable to the in-plane stiffness of the $Cu_2Si$ monolayer (93 N $m^{-1}$), and higher than germanene (42 N $m^{-1}$), suggesting that monolayers have good mechanical properties. For o-FeC monolayer, the calculated in-plane stiffness is 203 (along **a** direction) and 218 (along **b** direction), respectively. As these values are not equal to each other, o-FeC monolayer is mechanically anisotropic, and the in-plane stiffness of o-FeC is obviously higher than t-FeC. Thus, the proposed monolayers show strong mechanical stability.

In view of the unique bonding features in the monolayers, we then analyze the electron localization function (ELF) and deformation electron density (DED) to better understand the bonding features of these monolayers. The ELF analysis is a useful strategy for identifying and visualizing electron localization in molecules or solids.[44] The values are renormalized between 0.00 and 1.00, and in general, the value of 1.00 and 0.50 represents the fully localized and fully delocalized of the electrons, respectively; while the nearer values to zero, the lower charge density is.[45] As shown in Figure 3, the t-FeC monolayer exhibits a clearly ELF image: the ptC containing Fe-C network, where ELF distributes around the C-centered four Fe-C bonds. In t-FeC monolayer, DED map reveals that electron transfers from the Fe to C atom as shown in Figure 3c, and Bader analysis shows a charge transfer of 0.77 $e$ from each Fe atom. The transferred electrons, particularly from the Fe-d state, are delocalized around the four Fe-C bonds. Simultaneously, the C-p state is found to partially deplete, and delocalize over the four Fe-C bonds. Therefore, the delocalized electron states in t-FeC monolayer not only recede the atomic activity in forming out-of-plane bonds, but also strengthen the in-plane Fe-C bonds, which is crucial for stabilizing this planar four-coordinate moiety.[23,35]

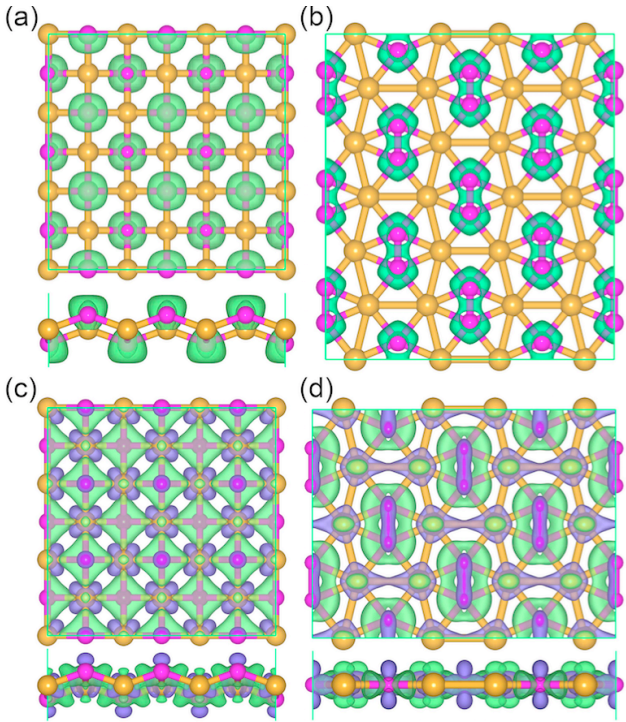

**Figure 3**. Iso-surfaces of ELF with the value of 0.70 for (a) t-FeC and (b) o-FeC, respectively. Deformation charge density of (c) t-FeC and (d) o-FeC monolayer, respectively. Green and blue refer to electron accumulation and depletion regions, respectively. The iso-value is 0.01 $e/Å^3$

Interestingly, the o-FeC monolayer exhibits dumbbell-like ELF image, which is featured by two C atoms surrounded by adjacent Fe atoms. In this novel ELF maps, one part is distributed around the Fe-C bonds, where ELF distributes around the C-centered Fe-C bonds; while another part is the C-C dimer structure. Therefore, o-FeC contains one C-C bond and four Fe-C bonds for one C atom. Bader analysis shows a charge transfer of 0.69 $e$ from each Fe atom and each C atom captures approximately 0.65 $e$ from Fe atoms. Moreover, as shown in Figure 3d, a significant electron transfer from Fe to C atoms, as identified by the different colored regions in the DED map, results in the electronic supplement for monolayer and thus stabilizes this pure planar framework. This significant electron transfer agrees with the larger electronegativity of Carbon (2.55) than Iron (1.83).

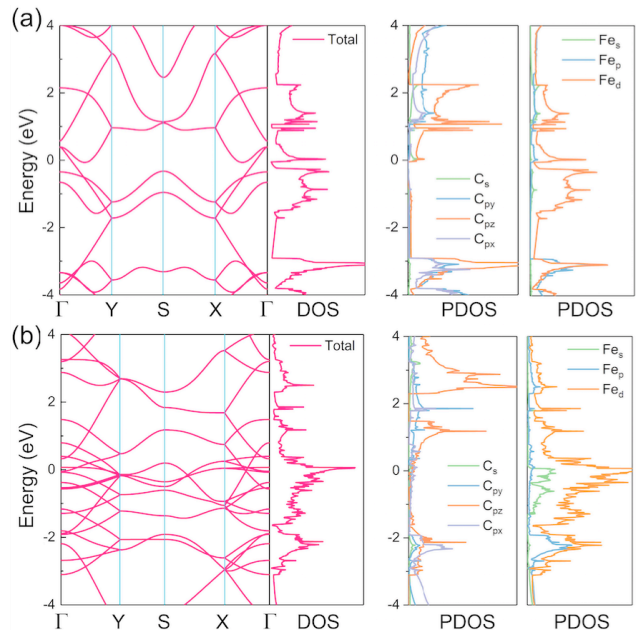

**Figure 4**. Band structure (left) and projected density of states (PDOS) of (a) t-FeC and (b) o-FeC, respectively. The Fermi level is assigned at 0 eV.

To get more insight into the electronic properties, we calculated the band structure and density of states (DOS) of the proposed monolayers. We also recalculated their band structures by using the GGA+$U$ scheme with $U$ = 5.0 eV for Fe-d orbital,[28] which are similar to the results achieved by using the GGA method (Figure S2). As shown in Figure 4, the metallic characters of monolayers are demonstrated by the Fermi level being located inside the bands, and no apparent the band gap around the Fermi level is observed. As expected, the strongest contributions at the Fermi level origin from the Fe-d and C-p states. For o-FeC, it is noteworthy that the high peaks of DOS appear around the Fermi level, suggesting the high density of carriers at the Fermi level. These high densities of electron states suggest the electrons that can efficiently participate the electronic transport propose, leading to the outstanding electron conductivity of the o-FeC monolayer. The accompanying electric conductivity is in accordance with the delocalized electrons as investigated by ELF and DED analysis.

As mentioned in the aforementioned section, the intrinsic metallicity of the global minimum t-FeC monolayer makes it a potential anode material in Li-ion batteries. Therefore, we investigated the adsorption energy and diffusion of Li-ion on the t-FeC monolayer. We first collected some high-symmetry adsorption sites, and then deposited a single Li atom on each adsorption site of a 3 × 3 supercell. As shown in Figure S3, the most favorable Li adsorption position located on the top of the C atoms, and the adsorption energy is -1.89 eV/Li-atom, suggesting the high stability of the t-FeC-Li complex systems.

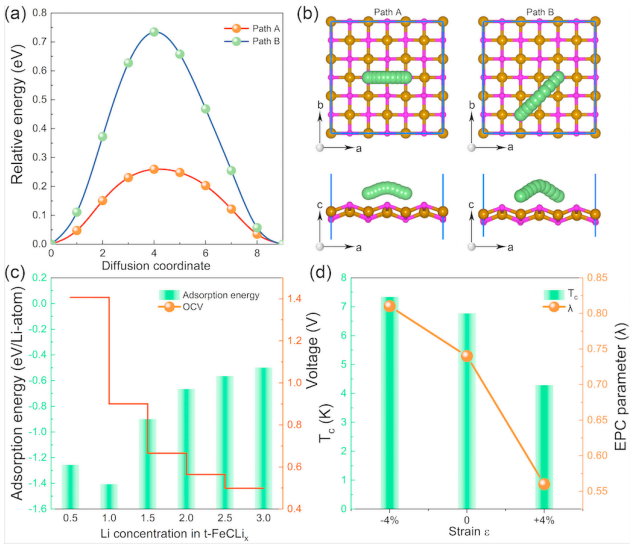

**Figure 5**. (a) Lithium diffusion pathway on the surface of t-FeC monolayer. (b) The corresponding diffusion barrier profiles of Li on t-FeC monolayer through the predesigned pathways (Path A and Path B). (c) The variation of adsorption energy and calculated voltage profile as a function of Li concentration in t-FeCLi$_x$. (d) and T$_c$ and EPC strength λ of t-FeC as functions of strains.

Two possible diffusion pathways are investigated in a 3 × 3 supercell as shown in Figure 5. When Li moves through the pathway A, only a small energy barrier of 0.26 eV should be overcome, which is close to that of MoS$_2$ (0.25 eV) and commercialized graphite anode (0.22 eV).[46] With the increase of Li-ion concentration, as shown in Figure 3c, the Li adsorption energy of t-FeCLi increases gradually with the increase of $x$. Because with increasing Li concentration, the distance between neighboring Li atoms becomes smaller, leading to more pronounced repulsive Coulomb interactions between adjacent Li atoms. Surprisingly, t-FeC monolayer can provide a Li adsorption energy of -0.50 eV/atom even at $x$ = 3, indicating that Li atoms can be stably adsorbed on monolayer at such a high concentration. Therefore, when t-FeC monolayer reaches a highest Li storage capacity, $x$ = 3 (t-FeCLi$_3$), the estimated OCV is 0.50 V, and the theoretical capacity is calculated to be 1184 mAh g$^{-1}$, which is higher than that of graphite and MoS$_2$ monolayer, demonstrating t-FeC is a promising candidate anode for Li-ion batteries.

The metallic character and existing strong bonding between C-Fe atoms in t-FeC monolayer also inspire us to investigate its electron-phonon coupling (EPC) and potential superconducting property. We then performed comprehensively EPC calculations for t-FeC monolayer to probe its superconductivity. An intriguing picture emerges when studying the dependence of the T$_c$ and λ on the uniaxial strain. Result for uniaxial strain applied in t-FeC is shown in Figure 5d. At zero strain, the calculated EPC parameter λ is 0.74, which suggests that the EPC is strong, and the phonon frequency logarithmic average ω$_{log}$ calculated directly from the phonon spectrum is 172 K. Therefore, using $\mu^* $ = 0.10-0.13, critical temperature T$_c$ for the t-FeC structure, estimated from the Allen-Dynes modified McMillan equation,[34] is in the range of 6.77-5.31 K, which is higher than previous theoretical or experimental reported some 2D materials, such as Y$_2$C (≈ 0.9 K),[47] Cubine (≈ 1 K),[48] and Mo$_2$C (≈ 3.4 K).[49] Moreover, previous literature demonstrated that external strain may greatly influence the superconducting property.[50,51] In our work, when a small uniaxial compression strain is applied (4%), the T$_c$ shows moderately increase (7.34 K), and λ becomes higher (0.81), demonstrating a strong EPC in t-FeC.

## 4. CONCLUSION

To summarize, by means of systematic *ab initio* computations, the energetically stable 2D Fe-C allotropes featured by planar hypercoordinate chemical bonding are proposed. In t- FeC monolayer, each C atom bonds to four Fe atoms to form a quasi-phC moiety. The predicted monolayers in this work are absence of imaginary phonon modes, and can be stable up to at least 1500 K in the FPMD simulations, indicating a great potential to be realized experimentally. The calculated strong adsorption energy, high theoretical capacity, and small diffusion barrier indicate that t-FeC is a promising candidate for Li-ion batteries. We also identified a theoretical prediction of superconductivity in t-FeC. Overall, unique structural and electronic features endow these 2D Fe-C allotropes with potential applications in future nano-devices.

ASSOCIATED CONTENT
**Supporting Information**
The *Supporting Information* is available free of charge on the...

AUTHOR INFORMATION
Corresponding Authors

*E-mail: lsh@zjut.edu.cn (Shaohua Lu)

*E-mail: huxj@zjut.edu.cn (Xiaojun Hu)

**Notes**
The authors declare no competing financial interest.

ACKNOWLEDGMENTS

This work was supported by the National Natural Science Foundation of China (Grant Nos. 11504325, 50972129, and 50602039), and Natural Science Foundation of Zhejiang Province (LQ15A040004). This work was also supported by the international science technology cooperation program of China (2014DFR51160), the National Key Research and Development Program of China (No. 2016YFE0133200), the European Union's Horizon 2020 Research and Innovation Staff Exchange (RISE) Scheme (No. 734578), and the One Belt and One Road International Cooperation Project from Key Research and Development Program of Zhejiang Province (2018C04021).